\documentclass[12pt]{article}
\usepackage[margin=1.5in]{geometry}                % See geometry.pdf to learn the layout options. There are lots.
\pdfoutput=1
\usepackage{graphicx}
\usepackage{amsmath}
\usepackage{amssymb}
\usepackage{fancyheadings}

\title{Ken Wilson --- The Early Years\footnote{\underline{Ken Wilson Memorial Volume}, World Scientific, Singapore}}
\author{R. Jackiw\\
\it \small Department of Physics\\
\it \small Massachusetts Institute of Technology\\
\it \small Cambridge, MA 02139}
\date{}                                           % Activate to display a given date or no date

\begin{document}
\maketitle
%\section{}
%\subsection{}
\thispagestyle{fancy}

Ken Wilson's life-time achievements in fundamental theoretical physics are well known 
and are well documented in this memorial volume. Therefore, there is no purpose in my 
adding yet another appreciation of his seminal work. Rather, I shall describe Ken and some 
of his activities at the beginning of his career, when he was junior faculty at Cornell and I 
was his student --- one of two in the first cohort of PhDÕs that he mentored. (The other was
Gerald Estberg, long time faculty and now retired at the University of San Diego.)
 
I entered Cornell's physics PhD program, hoping to study with Hans Bethe. But he decided 
to leave elementary particle physics and remain with nuclear physics. Another eminent 
theorist, specializing in S-matrix/Regge theory, left Ithaca for the West Coast. 
Consequently a position was offered to Ken, and in 1963 he accepted, partially ``because 
Cornell was a good university, was out in the country and [had] a good folk dancing 
group."\cite{jac1}.We graduate students were not familiar with his work because none was 
published.
Evidently he got the job solely on his reputation among senior colleagues as a brilliantly 
unique quantum field theorist. Perhaps this disappointed some, who wished to follow the 
then-dominant S-matrix approach. But I was delighted, because my ambition was to master 
quantum field theory.

We were bemused by Ken's dedicated work habits: One could find him in office most of the 
time; otherwise he resided in a motel room. We were again bemused two years later when 
he won tenure after two publications. He attended our parties and other informal 
gatherings. His interactions were marked by very deliberate responses to conversational 
gambits. One frequently had to wait some moments before he responded; when an answer 
came it was complete --- there was nothing more to say.

Ken's teaching style was methodical, addressing complicated matters in simple but 
opinionated fashion. He described his approach as
              \begin{quote}
               ``...not trying to state the final word on the physically (sic) meaning of quantum 
                fields. Rather, I ... present [an] intuitive understanding of the ... so-called
                `asymptotic condition'. Rather than go through the formal mathematics 
                involved in this work, I ... replace the rigorous but formal approach by an 
                intuitive hypothesis used in an intuitive way, to obtain the same results.
                The value ... is that it is obtained without ... introducing ideas which are 
                physically misleading and mathematically absurd. (`interaction
                representation' and the `adiabatic hypothesis')" \cite{jac2}
               \end{quote}
This attitude led to a clear but leisurely course presentation. By the end of the first 
semester of quantum field theory, we managed to quantize the free scalar field and discuss 
interactions, with no Feynman diagram in sight.                     

I presented myself to Ken and he agreed to direct my thesis research. He suggested that I 
study the renormalization group by reading the Gell-Mann Low paper \cite{jac3} and the 
Bogoliubov Shirkov \cite{jac4} text. Evidently already in 1963 Ken was thinking about the renormalization 
group. This choice came as a surprise to me because prevailing sentiment at that time 
maintained that nothing physically interesting can be gotten by renormalization group 
arguments, especially by techniques employed in the Soviet school. \cite{jac5}

             In fact Ken was an aficionado of the renormalization group from very 
             early days. Already in his 1961 PhD thesis, he used that formalism to solve the Low
             equation. The thesis also exhibits Ken's reliance on numerical, 
             computer assisted calculations --- another feature of his mature work. \cite{jac6}

When I was ready to begin research, Ken suggested that I use renormalization group 
methods to determine the large momentum behavior of the vertex (3-point) function in 
spinor electrodynamics. We hoped that rederiving known partial results would check the new 
approach, and that new results would demonstrate the power of the renormalization group 
in new settings. Let me explain.

The vertex function, depicted in the figure, describes the propagation of an off mass-shell 
electron (solid lines) with the emission of an off mass-shell photon (dashed line). The 4-momenta are, respectively $p,q$ and $k = p-q$. The on shell electron mass is $m$; the photon 
carries an infra-red regulator mass $\mu$.
%\begin{figure}[h!] %  figure placement: here, top, bottom, or page
%
%   \centering
%   \includegraphics[width=3in]{wilsonloops2.pdf} 
 %  \caption{Vistex (2-point) function}
%   \label{jackiwfig1}
%\end{figure}

$\Gamma^\alpha (p, q)$ is a 4 x 4 matrix, but the leading term may be isolated as $\Gamma^\alpha (p,q) = \gamma^\alpha \Gamma (p^2, q^2, k^2) + ....$ . The task is to study $\Gamma$  for large  $k^2$. The answer, far off mass-shell, $|k^2| \gg |p^2|, |q^2| \gg m^2$, was found by Sudakov.\cite{jac7}  
\[
\Gamma (p^2, q^2, k^2) \sim \exp   \left[ -\frac{\alpha}{2\pi} \ l n\  \left| \frac{p^2}{k^2} \right|\ ln\ \left|\frac{q^2}{k^2}\right|\ \right]\qquad \text{(Sudakov)}
\]

  \begin{center}
   \includegraphics[width=3in]{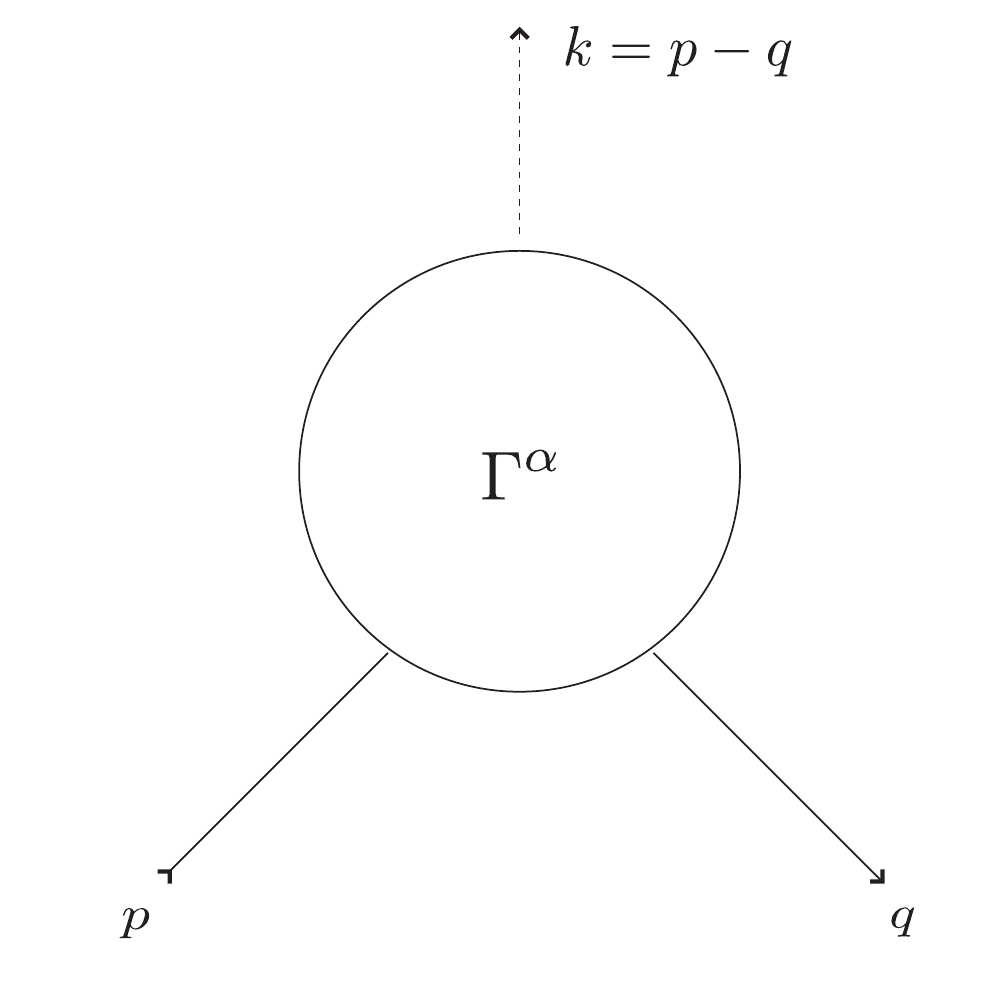}\\%[.5ex]
 Vertex (3-point) function
\end{center}

Rederiving this with the help of the renormalization group     
would validate that technique, and then  the on mass-shell formula for $\Gamma (m^2, m^2, k^2)$  could also be found.

Unfortunately I did not succeed. I could not find a defensible renormalization group 
argument for determining the large $k^2$ asymptote of  $\Gamma (p^2, q^2, k^2)$ off or on mass-shell. After some  futile struggle with the problem, I reported my failure to Ken.
I was afraid that he would lose interest, once the renormalization group was abandoned.

Fortunately he was open to other methods. After a few days he told me to take a different, 
eikonal-type approach: In a Feynman diagram expansion of $\Gamma^\alpha (p,q)$, a generic 
propagator should be decomposed as
\[
\begin{array}{rll}
  \frac{1}{r^2-M^2 + i \varepsilon}& = & \frac{1}{2 w_r} \left\{\frac{1}{r^0 - w_r + i\varepsilon} + \frac{1}{-r^0 - w_r + i\varepsilon}\right\}   \\[3ex]
 											 w_r &  = &  \sqrt{\mathbf{r}^2 + M^2}  \\ [.5ex]
  \multicolumn{3}{l}{\text{Decomposition of Feynman propagator}}
\end{array}
\]

\noindent Further analysis for large  momenta shows that only one of the two terms in the decomposition 
dominates. With this observation it becomes possible to sum all relevant graphs. The 
Sudakov formula is reproduced and the on mass-shell asymptote is found. \cite{jac8}
\[
\begin{array}{ccc}
&\Gamma (m^2, m^2, k^2) \sim \exp - \left\{\frac{\alpha}{4\pi} \ ln^2\ \frac{k^2}{\mu^2} \right\} \quad (\text{on mass-shell})&\\[1.5ex]
&m^2 = p^2, q^2 <\!\!<  | k^2 |&
\end{array}
\]

Note that the on mass-shell formula is not merely Sudakov's expression evaluated at $p^2 = q^2 =m^2$;
numerical factors differ. (With the advent of Effective Field Theory, a combination of 
eikonal and renormalization group methods can achieve both  on shell and off shell 
results. \cite{jac9})

While I was completing my research, there appeared a paper by S. Weinberg, \cite{jac10} in which 
he proposed  modified Feynman rules for calculating  amplitudes in the infinite 
momentum frame. These are very similar to Ken's suggestion. When I went to inform Ken, 
he had already seen the paper, and with a smile called my attention to it. Never did he 
claim any priority in this matter --- the subject simply was outside his interest, yet he could 
make a crucial contribution.

Ken was very supportive in my career. Upon my graduation he (and Bethe) secured for me a Harvard 
Junior Fellowship.  It gave me great pleasure that he too held one, just before me. One time  
when I saw him, he was revising a paper on his short distance expansion --- a technique with 
which he hoped to analyze the behavior of quantum fields, but had not yet come to fruition. 
I asked him how long would he remain with the subject without  establishing useful results. 
He answered that he wouldn't give up for a decade. But he didn't have to wait that long. In 
1969 he published the first of his renowned papers, ``Nonlagrangian Models of Current 
Algebra." In it he announced a new approach to quantum field theory. 
\begin{quote}
 ``What is proposed here is a new language for describing
the short-distance behavior of fields in strong interactions.
One talks about operator-product expansions
for products of operators near the same point, instead
of equal-time commutators. One discusses the dimension
of an operator instead of how it is formed from
products of canonical fields. Analyses of divergences in
radiative corrections, etc., are carried out in position
space rather than momentum space. Furthermore, one
has qualitative rules for the strength of SU(3) X SU(3)-symmetry-violating corrections at short distances...the hypotheses of this paper have the
elegance of simplicity, once one is used to the language." \cite{jac11}
\end{quote}
 \newpage
 \newgeometry{bottom=1.521in}
  
He worked out several illustrative examples of physical processes by his methods. Most 
pleasing to me is the fact that he devoted a long discussion to the chiral anomaly, which 
appeared the same year. \cite{jac12}

The Boston Joint Theoretical Physics Seminar met on Wednesdays. The day before 
Thanksgiving might have been sparsely attended; but this was not so, because the perennial speaker was Ken, who 
would be coming to visit his Boston area family for the holiday. As the years progressed, 
the audiences grew larger and larger,  and the largest room had to be used.

\restoregeometry
\end{document}